\documentclass[12pt]{article}
\usepackage{psfig}
\usepackage{graphicx}

\begin{document}

{Kinematika i Fizika Nebesnykh Tel, vol. 22, no. 6, pp. 422-431 (2006)} \\ \\ \\ \\

\begin{center}

\bigskip
{\large \bf Galaxy Triplets of the Local Supercluster. \\ 3. Configuration Properties} \\

{\small O. V. Melnyk$^{1}$, I. B. Vavilova$^{2}$}  \\

{\em Astronomical Observatory, Kiev National University, 3 Observatorna str., \\
Kiev, 04053 Ukraine \\
melnykov@observ.univ.kiev.ua \\
$^{2}$Space Research Institute. National Space Agency of Ukraine. \\ National Academy of Sciences of Ukraine.
40 Akdemika Glushkova av., Kiev, 03680 Ukraine
}\\
\end{center}

{\large \bf Abstract}\\
Configuration properties of observed galaxy triplets are studied with the use of the method of configuration parameters and Agekian-Anosova configuration diagram. Statistically significant excess of the hierarchical configurations for the galaxy triplets of the Local Supercluster (LS) is established. The samples of the Interacting, Northern and Southern galaxy triplets do not demonstrate the same excess. Our results testify that the model of the dark matter concentrated in the halo of individual galaxies is convenient for dynamically young systems to which the most of LS triplets belong, whereas the model of the dark matter concentrated in the triplet's volume is more convenient for the compact triplets. 


\bigskip

\section{Introduction}

In this paper, which is a continuation of our study of the properties of Local Superclusler (LS) galaxy triplets [4, 10], we analyze the configuration properties of several observed samples of galaxy triplets. We compare them with one another and with the data obtained by constructing dynamical models of galaxy trip­lets [5, 11, 13, 17, 25] and using the dependences of kinematical and virial characteristics of triplets on the types of their configurations. In addition to the sample of LS triplets [4], we also consider the combined sample of Northern [9] and Southern [8] triple systems and the sample of Interacting triplets from Vorontsov-Velyaminov's catalog [29] with a root-mean-square (rms) velocity of the galaxies in the group $S_{v}<$ 300 km s$^{-1}$. We give the latter samples as comparison samples, since the Northern and Southern triplets are the only lists of isolated triplets over the whole sky and the Interacting triplets are prominent represen­tatives of the closest systems (some of their properties are considered in [15]).

The configuration of a galaxy triple, i.e., the shape of the triangle formed by the galaxies, is an important characteristic of its dynamical evolution. Computer simulations of the dynamical evolution of triple systems show that a triple spends more than half of its lifetime in a state when two galaxies are close to one another, while the third galaxy is far from them [5, 16, 25, 28]. This triplet configuration is called hierarchical. How­ever, the pattern of the dark matter distribution is of great importance in the evolutionary processes of trip­lets. The presence of dark matter in the volume of a system accelerates the growth of a dynamical insta­bility in it and enhanced the system's chaotization, thereby suppressing the formation of long-lived close pairs [5, 13].

The configuration characteristics of observed and simulated galaxy triplets have been the subject of many sludies [3, 5, 12, 16, 18, 19, 30], which revealed no excess of hierarchical configurations in the observed triplets. The samples of 46 triplets from the list of Northern triple systems [9] and 37 Wide triplets [16] from the catalogs [22, 26] were used as the observational material. Only the systems that, according to Anosova's statistical criterion [2], had the status of physical systems were selected.

Our study is aimed not only at ascertaining the pattern of the dark matter distribution in LS triplets, but also at augmenting the statistics of configuration properties of observed galaxy triplet samples.

\section{GALAXY TRIPLET CONFIGURATIONS}

To present the configuration characteristics of galaxy triplets, we chose the convenient and illustrative method of Agekyan and Anosova [1], which was widely used in [3, 5, 16, 18, 19, 30]: if the sides of the triangle are normalized in such a way that the galaxies forming the largest side have coordinates (-0.5, 0) and (0.5, 0) and the third galaxy lies inside the configuration triangle, then the triplet configuration (its coor­dinates) is uniquely determined by a pair of numbers $x\ge0$ and $y\ge0$, $(x + 0.5)^{2} + y^{2} \le 1$. If $ (x-0.5)^{2} + y^{2} \le 1/9$, then the triplet configuration is called an hierarchical one (H); if  $(x - 0.5)^{2} + y^{2} < 4/9$ and $(x - 0.5)^{2} +y^{2} > 1/9$, $y > -0.5x + 1/4$, then the configuration is called a medium one (M); if $(x-0.5)^{2} + y^{2} \ge 4/9$, then the configuration is close to an equilateral tri­angle and is called a Lagrangian one (L); if $(x - 0.5)^{2} + y^{2} < 4/9$ and $(x - 0.5)^{2} + y^{2} > 1/9$ and $y \le -0.5x + 1/4$, then the triplet configuration is called a linear one (A).

Figure 1 presents the configuration tri­angles for the triplet samples under con­sideration. The Northern and Southern triplets are clearly seen to be distributed more uniformly over the diagram than the samples of LS and Interacting triplets. Figure 2 presents the distributions of trip­lets in configuration types, where the rela­tive galaxy density ${\alpha}_{i}=\bar{\rho}_{i}/\bar{\rho}$ ($\bar{\rho}$ is the mean galaxy density in the entire configu­ration triangle and $\bar{\rho}_{i}$ is the mean galaxy density in the ith zone H, M, L, A) is along the $y$ axis. We compared the observed rel­ative galaxy density in each zone with the simulated, randomly specified triplets by taking inio account the projection effect (the dashed line in Fig. 2) by randomly specifying three points in a circle of unit radius and projecting them onto a ran­domly oriented plane of the sky. As a result, we obtained 20000 simulated cata­logs, each with $N_{tr}$ triplets in the corre­sponding sample, for estimating the num­ber of triplets of different configuration types (H, M, L, A) in each of these cata­logs and estimating the mean and standard deviation among the catalogs.

\begin{figure}[t]
\begin{tabular}{ll}
(a) & (b) \\\\
\includegraphics[angle=0, width=0.40\textwidth]{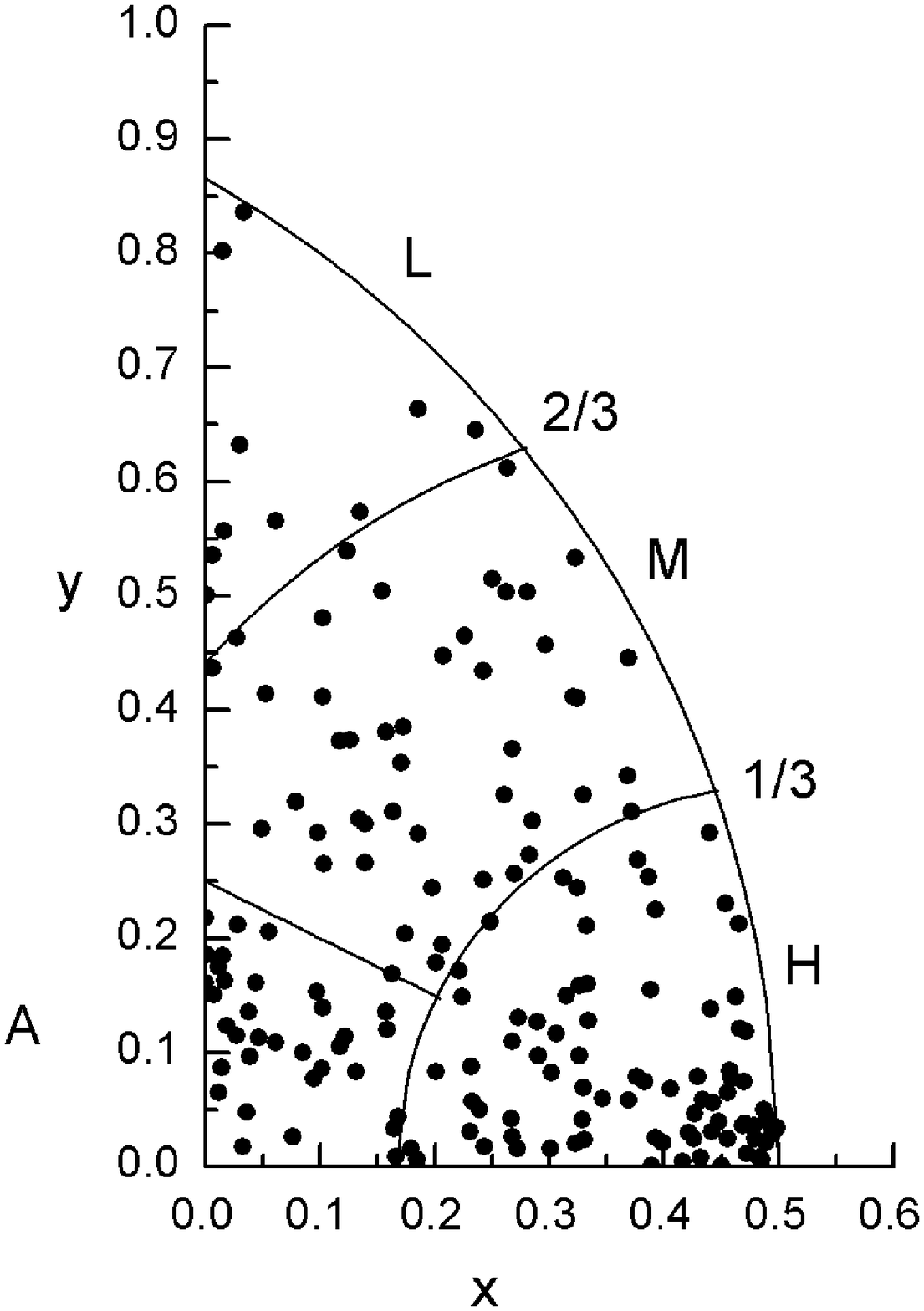} &
\includegraphics[angle=0, width=0.40\textwidth]{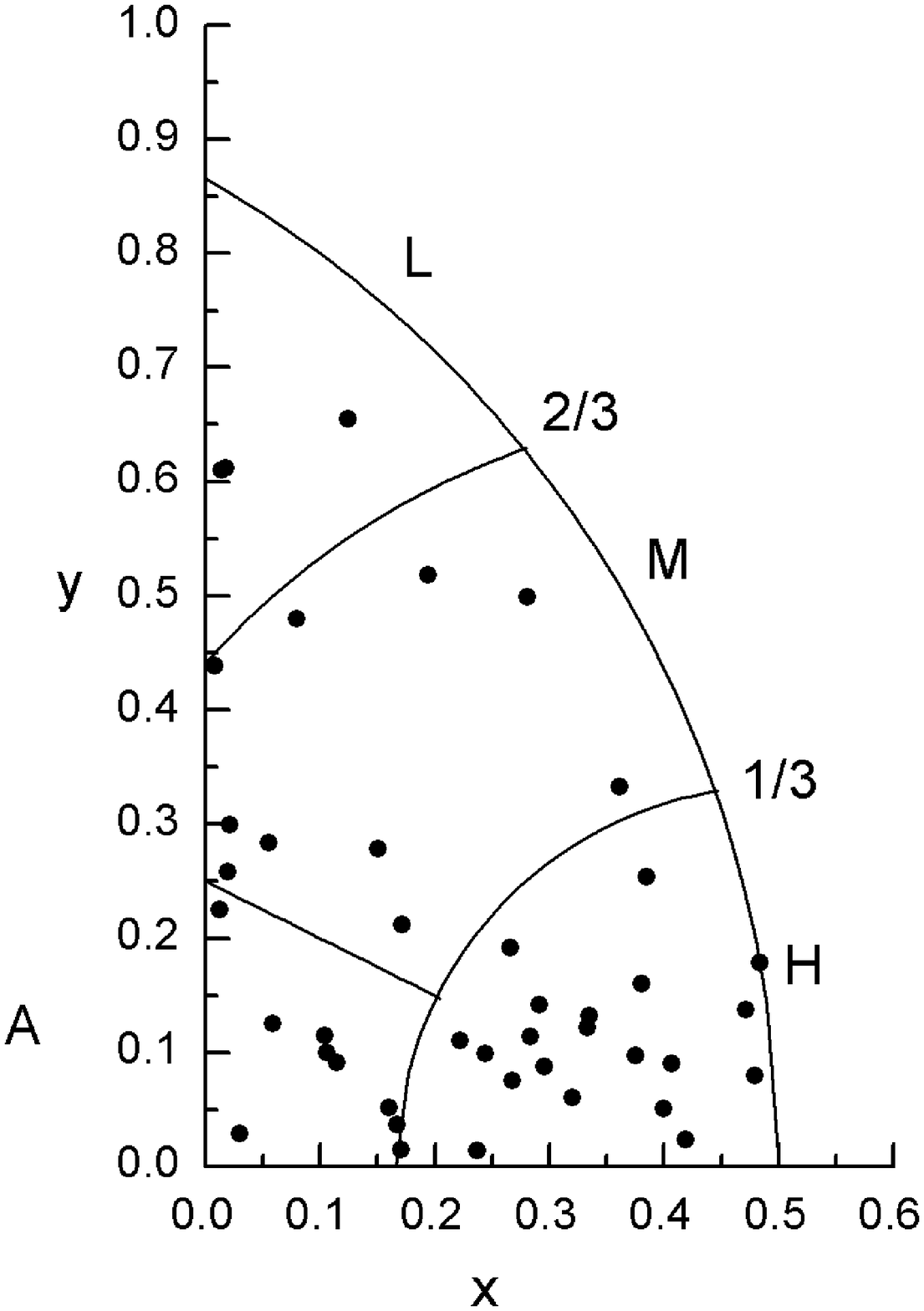} \\
(c) & (d) \\\\
\includegraphics[angle=0, width=0.40\textwidth]{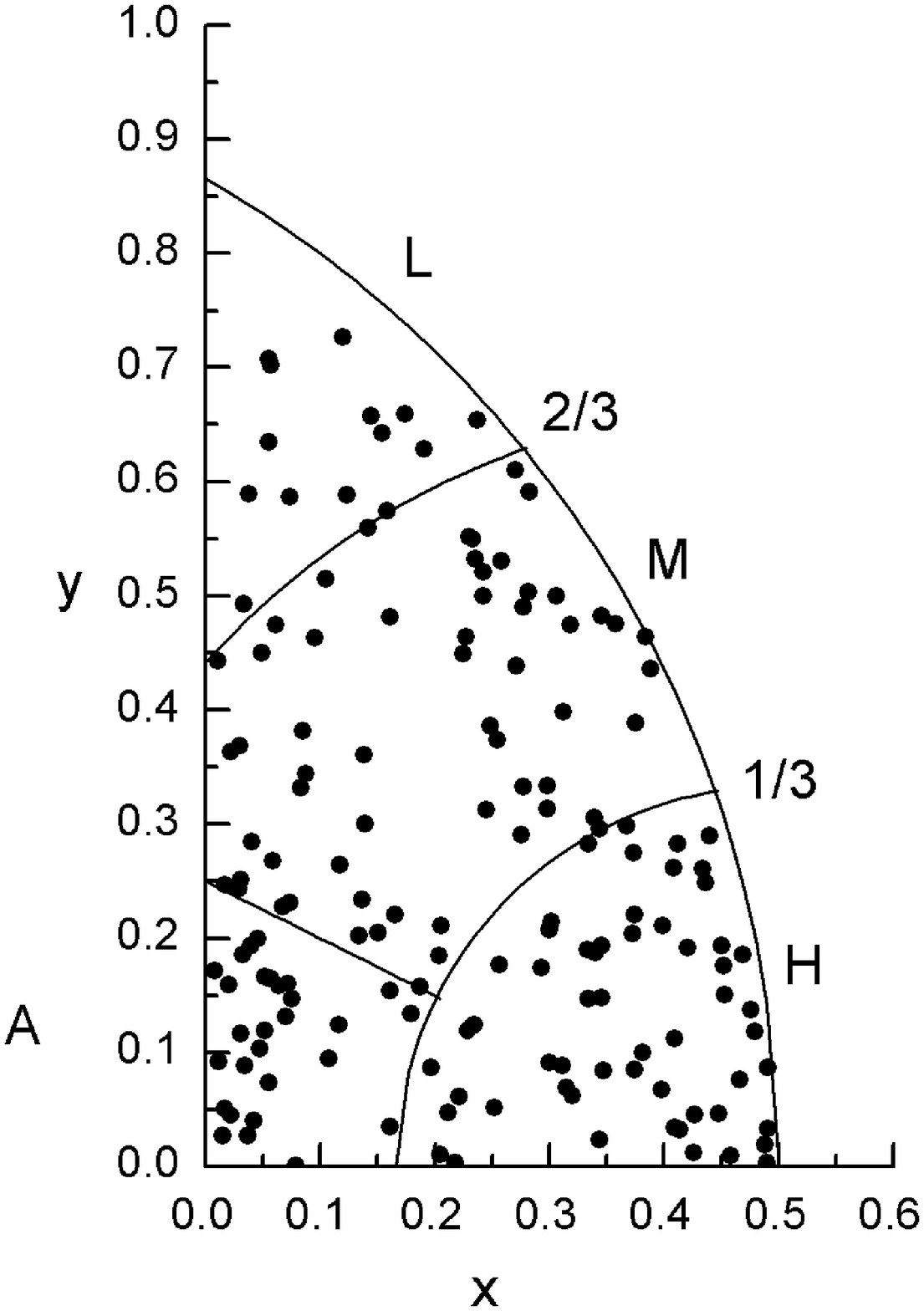} &
\includegraphics[angle=0, width=0.40\textwidth]{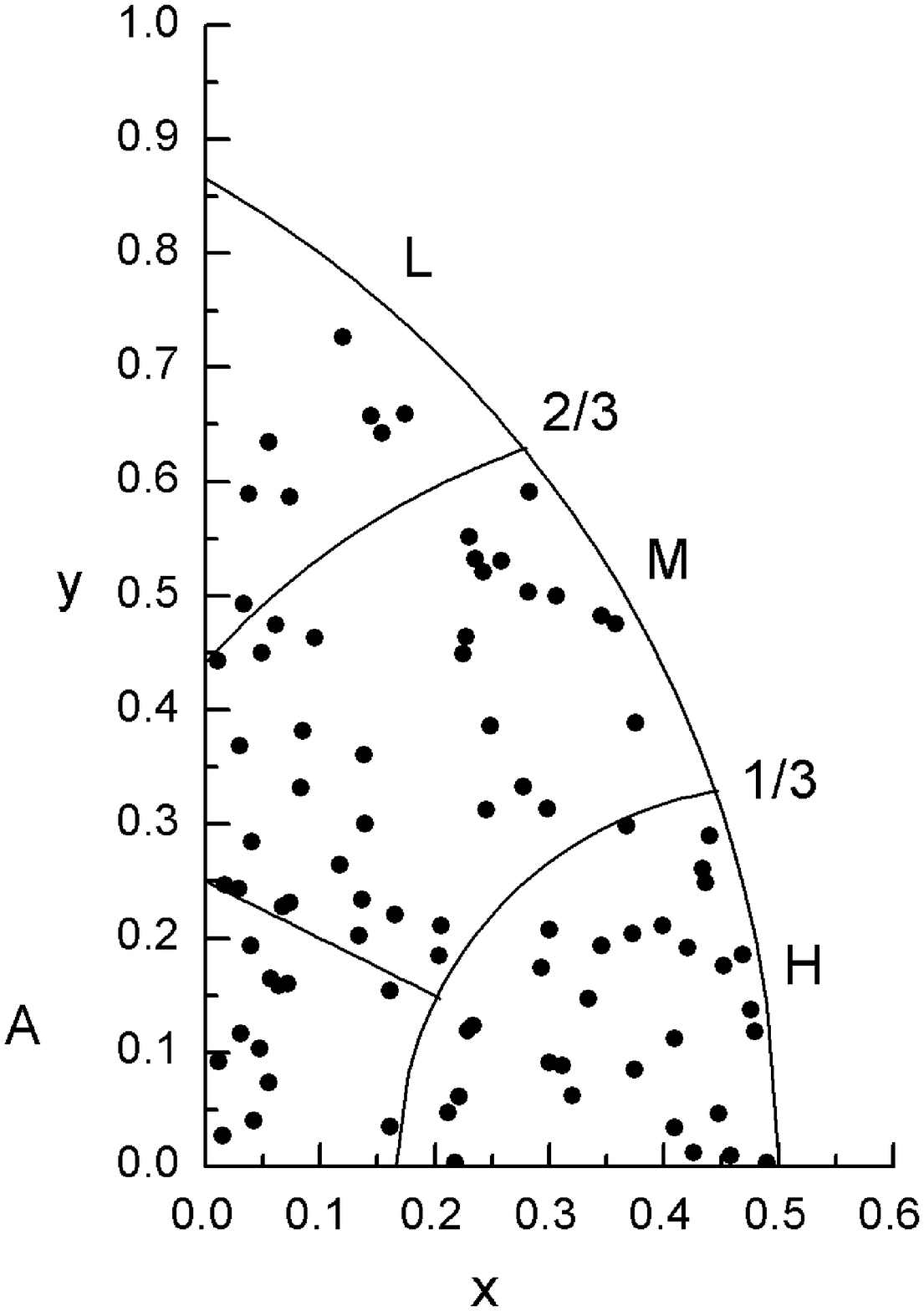} \\

\end{tabular}
\caption{Configuration diagram with a division into zones: (a) LS triplets, (b) Interacting triplets, (c) Northern and Southern triplets, and (d) Northern and Southern triplets with $S_{v} <$ 300 km s$^{-1}$. Regions H, L, A, and M are hierarchical, Langrangian, linear, and medium (intermediate) configurations, respectively.}
\end{figure}

We can see from Fig. 2 that the largest excess of hierarchical configurations is observed in the LS sample (the deviation from the mean is 2.9${\sigma}$). The samples pre­sented in Figs. 2c and 2d, the full com­bined catalog of Northern and Southern triplets and the presumed physical Northern and Southern triplets with $S_{v}< 300$ km s$^{-1}$, show no statisti­cally significant differences in the configurations both between themselves and with the random uniform distribution. Below, by the sample of Northern and Southern triplets we will mean only the systems with $S_{v} < 300$ km s$^{-1}$ (the same constraint as that for the interacting systems). Since the sample of Northern and Southern triplets under consideration and the sample of triplets considered previously [3, 5, 12, 13, 17-19, 30] are subsamples of the general list of Northern and Southern triplets [8, 9], the triplet systems in them should overlap.

To independently estimate the hierarchy and elongation of triple systems, we also used the method of configuration parameters [3, 12]. We calculated the hierarchy, ${\lambda}$, and elongation, ${\beta}$, parameters for the samples of observed triplets under consideration, which least depend to the projection effect [3, 12]:

\begin{equation}
\label{trivial}
\lambda = 2r_{min}/(r_{int}+r_{max}), \beta=\pi-\varphi_{max},   
\end{equation}

where $r_{min}$, $r_{int}$ and $r_{max}$ are the smallest, intermediate, and largest sides in the triangle formed by three gal­axies on the celestial sphere, $\varphi_{max}$ is the largest angle in the triangle. In this case, ${\lambda} \in [0, 1]$, where ${\lambda}$ = 0 corresponds to the case where the coordinates of two galaxies coincide and the third galaxy lies at some distance from them, while ${\lambda}$=1 corresponds to the case of an equilateral triangle. The elongation parameter ${\beta} \in [0,2{\pi}/3]$, where ${\beta}$ = 0 corresponds to a linear configuration and  ${\beta} = 2{\pi}/3$  corresponds to an equilateral triangle. The mean values of the configuration parameters, ${\lambda}'$ and ${\beta}'$, for the observed samples for randomly specified 20000 catalogs of triplets, each with $N_{tr}$ triplets within a circle of unit radius, with allowance made for the projection effect, their standard deviations, ${\sigma}_{{\lambda}'}$ and  ${\sigma}_{{\beta}'}$, and the deviations from the mean in units of the rms deviation, $({\lambda}'-{\lambda})/{\sigma}_{{\lambda}'}$ and $({\beta}'-{\beta})/{\sigma}_{{\beta}'}$, are presented in Table 1. Analysis of our results leads us to conclude that the triplets from the LS sample show a strong tendency to hierarchy (at a 4.1${\sigma}$ level); the Inter­acting triples show a weak tendency to hierarchy. Both samples of LS and Interacting triplets also show a weak tendency to elongation, while the Northern and Southern triplets show a very weak tendency to a Lagrangian configuration.
 
\parskip=3 mm
Table 1. Mean values of the configuration parameters of the samples of the LS, Interacting (In) and Northern and Southern (NS) galaxy triplets

\begin{tabular}{|l|l|l|l|l|l|l|l|l|l|}
\hline
Triplet sample & $N_{tr}$ & ${\lambda}$ & ${\lambda}'$ & ${\sigma}_{{\lambda}'}$ & $\frac{{\lambda}'-{\lambda}}{{\sigma}_{{\lambda}'}}$ & ${\beta}$ & ${\beta}'$ & ${\sigma}_{{\beta}'}$ & $\frac{{\beta}'-{\beta}}{{\sigma}_{{\beta}'}}$\\
\hline
LS& 173 & 0.392 & 0.458 & 0.016 & 4.1 & 0.921 & 1.002 & 0.045 & 1.8\\
\hline
In &42 & 0.422 & 0.459 & 0.033 & 1.1 & 0.892 & 1.002 & 0.090 & 1.2\\
\hline
NS &86 & 0.474 & 0.458 & 0.023 & -0.7 & 1.108 & 1.001 & 0.063 & -1.7\\
\hline
\end{tabular}

\parskip=3 mm

Table 2 presents the median parameters of ihe triplet samples under consideration for various configura­tions. Here, $N_{tr}$ is the number of triplets with the corresponding configuration, $S_{v}$ (km s$^{-1}$) is the rms velocity of the galaxies in the group, $R_{h}$ (kpc) is the harmonic mean radius of the system, ${\tau}$ ($1/H_{0}$) is the characteristic system crossing time, $\log(M_{vir}/M_{\odot})$ is the logarithm of virial mass, $\log(L/L_{\odot})$ is the logarithm of the group's total luminosity, $(M_{vir}/L)/(M_{\odot}/L_{\odot})$ is the virial mass-to-light ratio, and $N(E/S0)/(3N_{tr})$ is the fraction of early-type galaxies in each subsample. All of the calculations were performed using formulas from [4]; the Hubble constant was assumed lo be $H_{0}$ = 75 km s$^{-1}$ kpc$^{-1}$. It can be seen from Table 2 that the LS triplets are young systems with a long crossing time, a low rms velocity of the galaxies in the group, and a large harmonic mean radius, while the triplets of the other two samples have a short crossing time, which may be indicate of their old age.

\begin{figure}[t]
\begin{tabular}{ll}
(a) & (b) \\\\
\includegraphics[angle=0, width=0.44\textwidth]{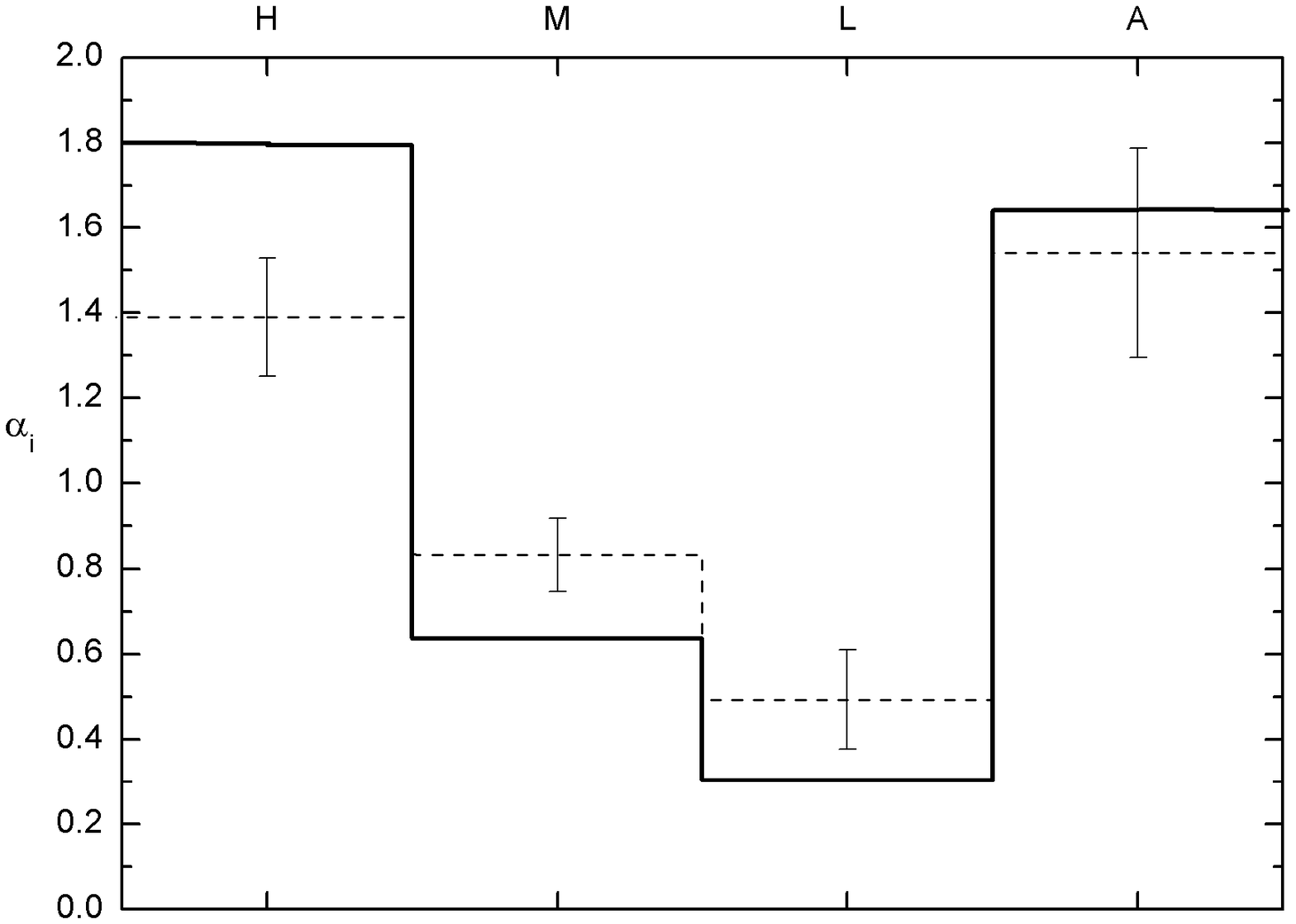} &
\includegraphics[angle=0, width=0.405\textwidth]{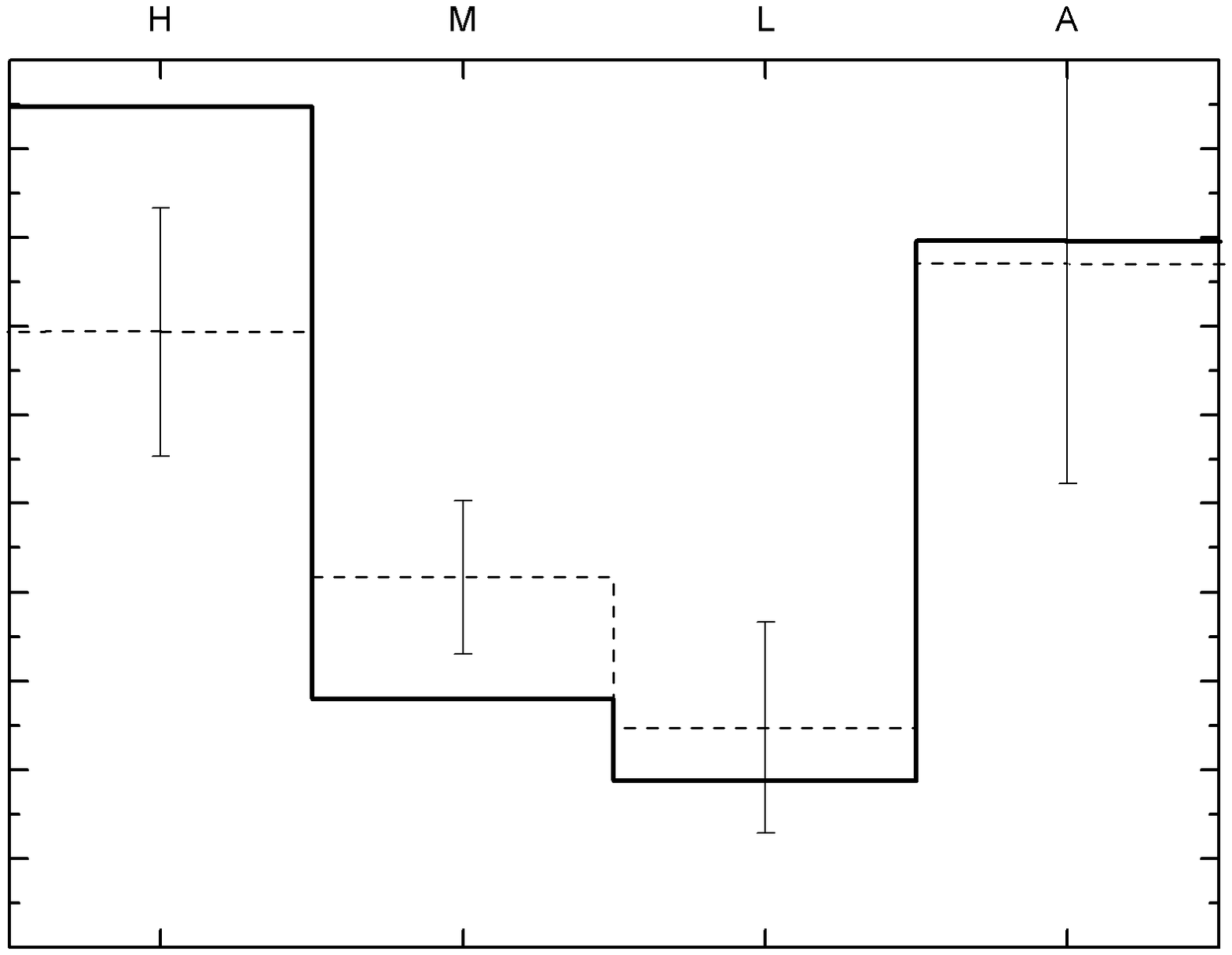} \\\\
(c) & (d) \\\\
\includegraphics[angle=0, width=0.44\textwidth]{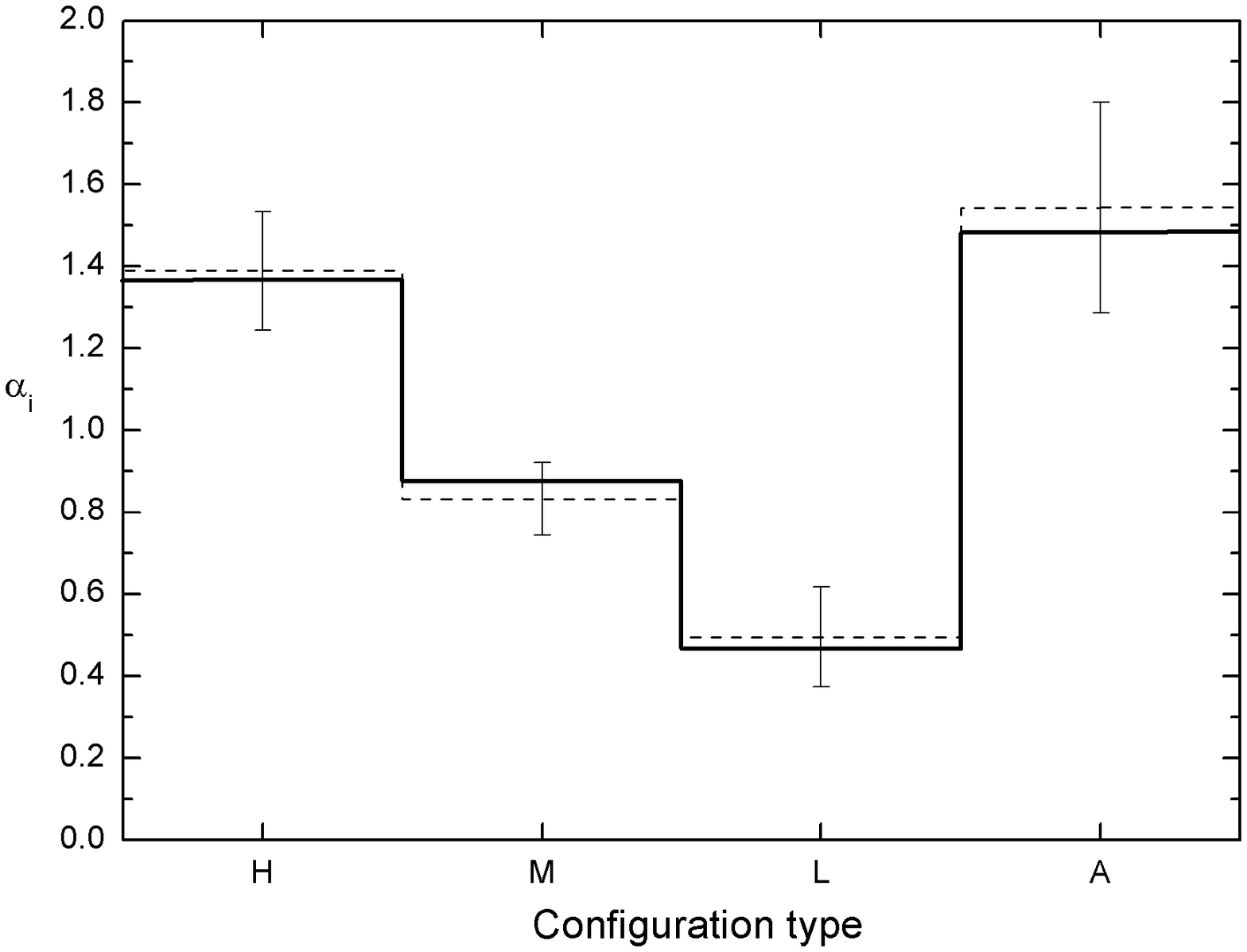} &
\includegraphics[angle=0, width=0.405\textwidth]{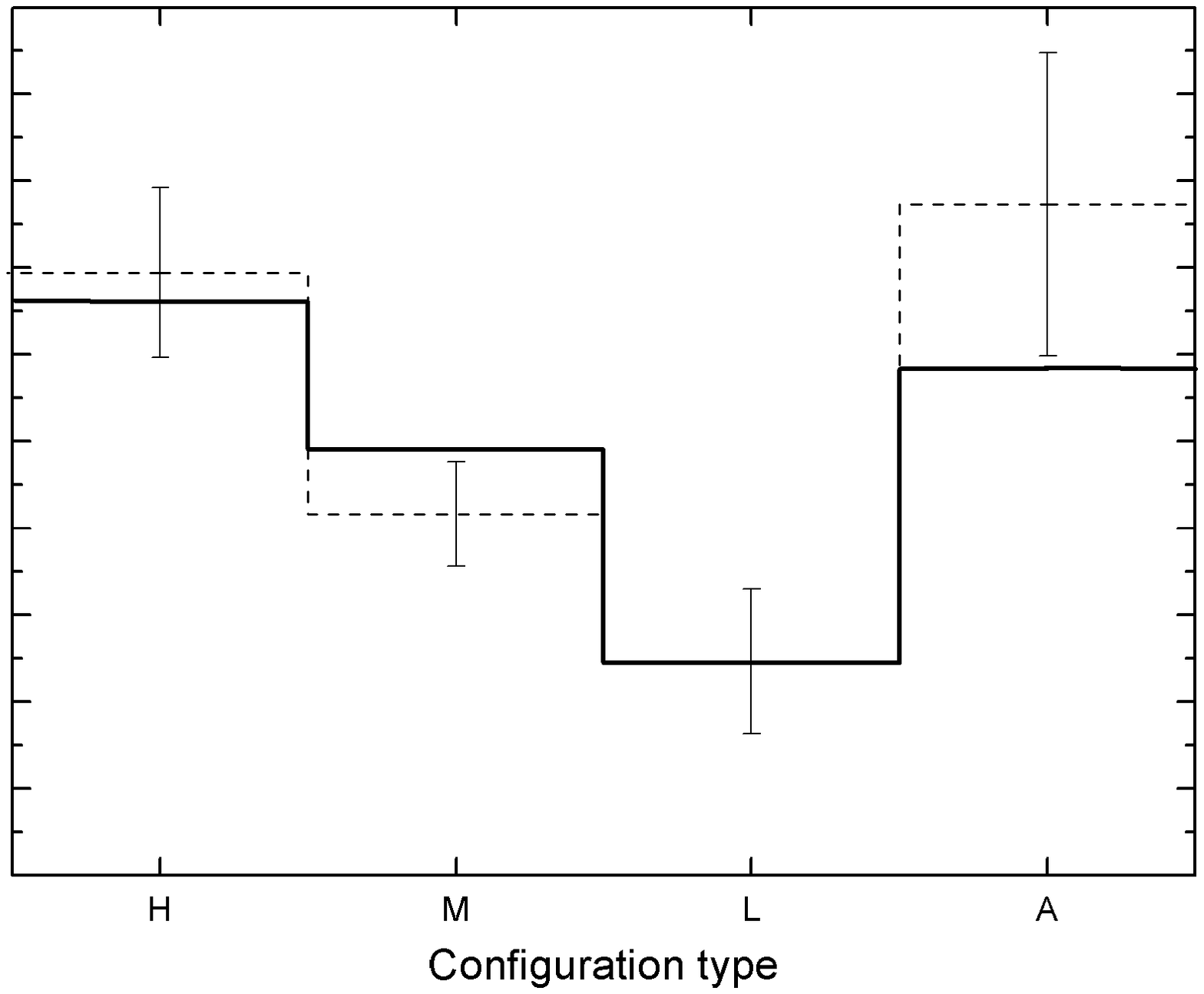} \\
\end{tabular}
\caption{Distribution of triplets in configuration types normalized to the area of the configuration region: (a) LS triplets, (b) Interacting triplets, (c) Northern and Southern triplets, and (d) Northern and Southern triplets with $S_{v} <$ 300 km s$^{-1}$. The dashed line represents a random uniform distribution with allowance made for the projection effect (see the text).}
\end{figure}

We see from Table 2 and Fig. 3 that the systems under consideration differ by the characteristic sepa­ration between the galaxies. The Interacting triplets are the closest systems of galaxies with a character­istic separation of the order of the galaxy size. The LS triplets are widest, while the Northern and South­ern   triplets   have   an   intermediate   value   of  $R_{h}$, although they are closer to the Interacting systems in all other characteristics. Despite these, as we believe, evolutionary differences between the systems under consideration, certain dependences on the configura­tion type (H, M, L, A) are observed for all samples. Such characteristics as the system crossing lime ${\tau}$, $\log(M_{vir}/M_{\odot})$, and $\log(L/L_{\odot})$ as well as the mass-to-light ratio are systematically smaller for the galaxy triplets from all of the samples considered with an hierarchical   configuration.  The   largest   values  of these characteristics are characteristic of the triplets with a Lagrangian and/or linear configuration. The Interacting  triplets  constitute  an  exception,  since very close chain-like systems represent a linear configuration here.

Table 2. Median of the kinematical and virial parameters for the LS, Interacting (In) and Northern and Southern (NS) galaxy triplets and the fraction of early-type galaxies in them as a function of the configuration type

\begin{tabular}{|l|l|l|l|l|l|l|l|l|}
\hline
Sample & $N_{tr}$ & $S_{v}$ & $R_{h}$ & ${\tau}$ & $\log(M_{vir}/M_{\odot})$ & $\log(L/L_{\odot})$ & $\frac{M_{vir}/L}{M_{\odot}/L_{\odot}}$ & $\frac{N(E/S0)}{3N_{tr}}$\\
\hline
\multicolumn{9}{|c|}{\textbf{H configuration}}\\
LS& 82 & 33 & 103& 0.45 & 11.47 & 10.24 & 24 & 0.21\\
In& 21 & 81 & 23& 0.04 & 11.56 & 10.76 & 14 & 0.30\\
NS& 24 & 95 & 50& 0.09 & 12.00 & 10.62 & 21 & 0.27\\
\multicolumn{9}{|c|}{\textbf{M configuration}}\\
LS& 47 & 26 &188& 0.94& 11.63 & 10.17 & 40 & 0.19\\
In& 10 & 97 & 30& 0.05 & 11.91 & 10.74 & 19 & 0.30\\
NS& 25 & 95 & 59 & 0.09 & 12.24 & 10.54 & 51 & 0.23\\
\multicolumn{9}{|c|}{\textbf{L configuration}}\\
LS& 10 & 22 &251& 2.36& 11.44 & 9.99 & 46 & 0.13\\
In& 3 & 108 & 44& 0.06 & 12.23 & 10.86 & 14 & 0.25\\
NS& 6 & 108 & 82 & 0.14 & 12.33 & 10.74 & 62 & 0.25\\
\multicolumn{9}{|c|}{\textbf{A configuration}}\\
LS& 34 & 27 &271& 1.81& 11.68 & 10.14 & 43 & 0.13\\
In& 6 & 84 & 25& 0.05 & 11.70 & 10.83 & 7 & 0.25\\
NS& 9 & 128 & 73 & 0.15 & 12.45 & 10.63 & 70 & 0.25\\
\hline
\end{tabular}

\parskip=3 mm
The dependence of the harmonic mean radius $R_{h}$ of a triplet on configuration parameter ${\lambda}$ gives the fol­lowing regression equations 
(where $R$ is the correlation coefficient and $SD$ is the standard (rms) deviation): $\log R_{h}$ = $0.58\log{\lambda}$ + 2.46, $R$ = 0.58, $SD$ = 
0.32 for the LS triplets; $\log R_{h}$ = $0.09\log{\lambda}$ + 1.45, $R$ = 0.32, $SD$ = 0.32 for the Interacting triplets; and 
$\log R_{h}$ = $0.51\log{\lambda}$ + 1.93, $R$ = 0.44, $SD$ = 0.35 for the Northern and Southern triplets.

Thus, the correlation and the dependence are strongest for the LS triplets, the dependence for the North­ern and Southern triplets is slightly weaker, and no dependence is observed for the close Interacting systems. Indeed, if the number of triplets in the sample is limited in $R_{h}$, then the smaller the radius         $R_{h}$ the more the hierarchical systems remain in the sample. However, such dependences are typical only for the samples of LS and Northern and Southern triplets. There is no such dependence for the Interacting triplets and the kinematic and virial characteristics of a triplet depend weakly on its configuration type. It may be assumed, according to Fig. 3. that, although the Northern and Southern triplets are, on average, compact, relatively wide systems are also encountered among them (a system was selected in the catalog not by the compact­ness criterion, as in the Interacting systems, but by the degree of isolation). Therefore, both wide systems similar in properties to the LS triplets and close systems similar in properties to the Interacting triplets can be encountered among the Northern and Southern triplets.

\begin{figure}[t]
\centerline{\includegraphics[angle=0, width=15 cm]{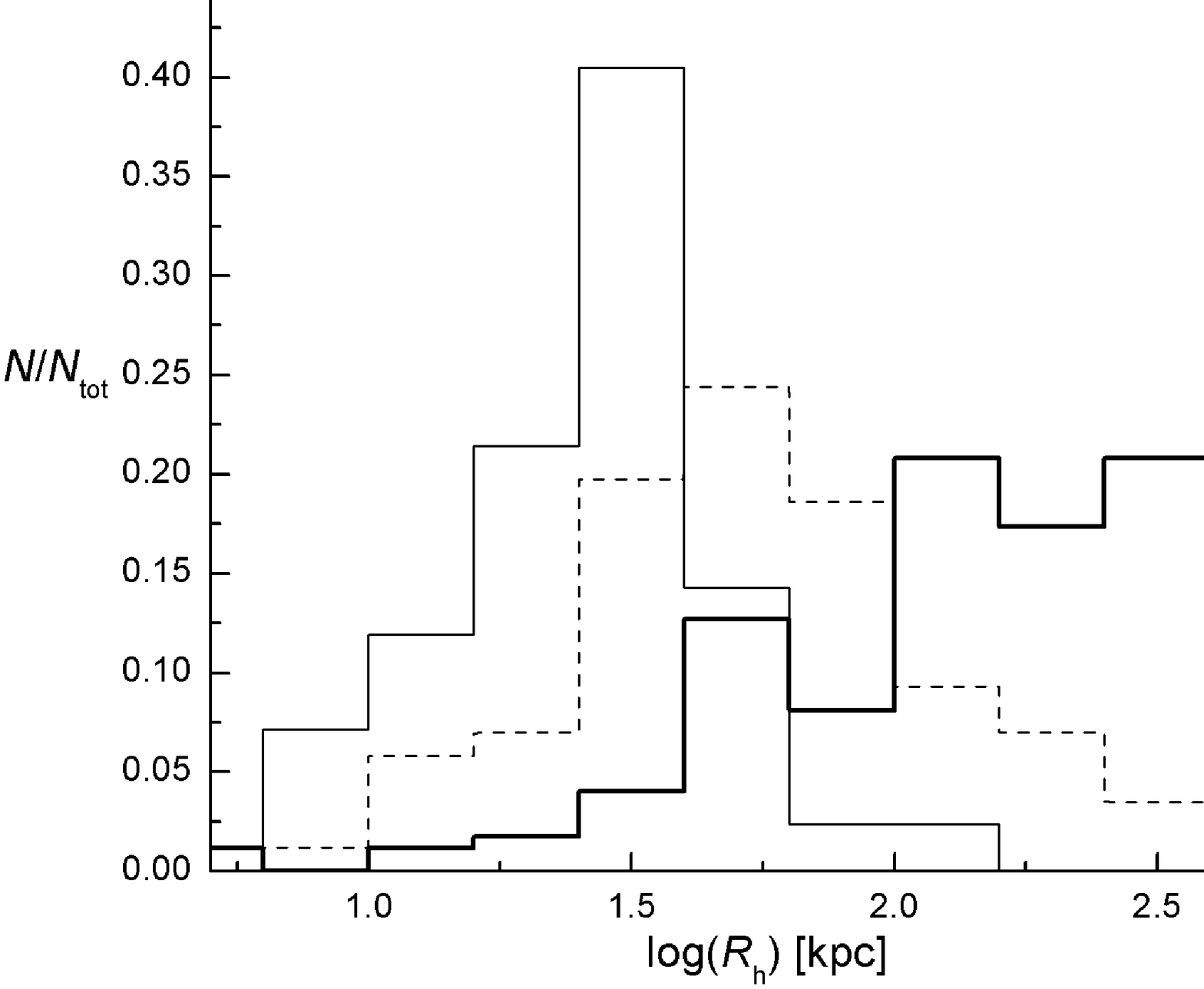}}
\caption{Distribution of galaxy triplets in harmonic mean radius $R_{h}$. The thick, thin, and dashed lines represent the LS sample, the Interacting triplets, and the Northern and Southern triplets, respectively.}
\end{figure}

A morphological analysis of the triplets considered shows (Table 2) that the fraction of early-type gal­axies in the LS triplets in medium and hierarchical configurations is larger than that in Lagrangian and linear configurations. The fraction of $E/S0$ galaxies for all LS triplets is 18 \%. There are 46 interacting galaxies from Vorontsov-Velyaminov's catalog [29] in 35 LS triplet systems; half of them also have an hierarchical configuration. Two triplets where all three galaxies interact have a Lagrangian configuration. These data can also be interpreted in a different way: of the 82 H triplets with a close pair, 11 pairs interact (13 \%).

Analysis of the distribution of triplet configurations in the LS volume shows that, depending on the mean radial velocity of the triplet, the number of triplets of various configurations exhibits no statistically signif­icant excess in the distribution for any configuration.

\section{DISCUSSION AND CONCLUSIONS}

We found differences in kinematical and virial characteristics of galaxy triplets from three samples of triplets, depending on the type of their configuration: H (hierarchical), M (medium), L (Lagrangian), and A (linear). Such characteristics as the system crossing time ${\tau}$, $\log(M_{vir}/M_{\odot})$, $\log(L/L_{\odot})$, and the mass-to-light ratio $(M_{vir}/L)/(M_{\odot}/L_{\odot})$ are systematically smaller than those for galaxy triplets with an hierarchical configu­ration for all of the samples considered. These characteristics are largest for triplets with a Lagrangian and/or linear configuration.

Using the Agekyan-Anosova configuration diagram [1] and the method of configuration parameters [3, 12], we established a statistically significant excess of hierarchical configurations in the sample of LS triplets (at 2.9${\sigma}$ and 4.1${\sigma}$ confidence levels, depending on the method, respectively). According to [14], an hierarchical configuration can form during the first several crossing times. Since the median crossing time is 0.87 for the entire ensemble of LS galaxy triplets and 0.45 for the hierarchical configurations, this suggests that 1-3 crossings have already occurred in the systems, which could give rise to hierarchical con­figurations, This result agrees with dynamical simulations [5, 17]. For example, Chernin et al. [17] found an excess of hierarchical configurations in the simulated triplets at 6.4${\sigma}$ and 7.3${\sigma}$ confidence levels, depend­ing on the presence or absence of hidden mass in the halo. When investigating an ensemble of triple systems with individual halos, Ivanov et al. [5] showed than the initial position of a triplet on the configuration dia­gram could not be traced even after 2.5 crossing times. They also concluded that hierarchical configurations dominate among other configuration types in the simulated triplets if dark matter is distributed in the halos of individual galaxies.

We compared real and random triplets by randomly specifying three points in a circle of unit radius and by projecting them onto a randomly oriented plane of the sky. Another method could be a comparison of observed triplets with a uniform distribution of the same number of points on the configuration diagram; the relative galaxy density ${\alpha}_{i}$, would then be equal to unity (Fig. 2), while the excess of hierarchical configura­tions would be even larger, i.e., the criterion we chose is more stringent. Of considerable importance is also the fact that we do not simulate the physical triplets. However, the simulations by other researchers [5, 17] allow conclusions about the distribution of dark matter to be reached by analyzing the distribution of con­figurations of a specific sample.

In turn, the samples of Interacting and Northern and Southern triplets show no statistically significant predominance of hierarchical configurations. Previously, a similar result was obtained for triple systems from the list of Northern triplets [3, 5,12,18,19, 30]. The model of dark matter distributed in the volume of a triplet could explain these data, since the H configurations of galaxy triplets become short-lived ones under these conditions [5, 13, 16, 19, 25]. Chernin el al. [17] showed that no excess of hierarchical config­urations was observed in the simulated triplets with dark matter concentrated in the common halo, as in the observed Northern triplets. In addition to theoretical models, observational data also provide evidenee for common halos of close groups. Extended X-ray halos were found in compact groups of galaxies, in partic­ular, in the triplet VV118 [24] from the sample of Interacting triplets and in the Northern and Southern trip­lets KTS48 = HCG67, KTS66 =HCG90, and KTG50 = HCG68 (since $S_{v}$= 410 km s$^{-1}$ for the latter triplet, it was not included in the resulting sample) [27]. Da Rocha et al. [21] found a diffuse stellar component of the halo in several compact groups in the optical range, but these observations are few in number and it is premature to talk about any dependences. However, there are assumptions of an evolutionary nature, for example, that the presence of a common (X-ray or diffuse) halo depends on the evolution level of the group, which can be characterized by the morphological composition of the group, the activity of galaxies in the group, and the separation between the components, the characteristic crossing time ${\tau}$ [19-21]. The Interact­ing, Northern, Southern triplets are at a later evolutionary phase than the LS galaxy triplets (${\tau}$ = 0.04, 0.09 and 0.87, respectively). Since the median value of the harmonic mean radius of the system in these two sam­ples is almost equal to the size of one galaxy (30-50 kpc), the individual and common halos of the triplet cannot be distinguished in such close systems.

The morphological composition of the galaxies in a group can serve as an indicator of whether there were mergers in the group in the past [19, 30]. There are 18 \% of early-type ($E/S0$) galaxies in the sample of LS trip­lets and 11 \% among the isolated LS galaxies (the LS sample is limited in radial velocity $V_{LG} <$ 3100 km s$^{-1}$). It was also found that the fraction of early-type galaxies in the LS triplets in medium and hierarchical con­figurations is larger than that in Lagrangian and linear configurations. Among the magnitude-limited sam­ples, there are 33 \% of early-type ($E/S0$) galaxies in the Interacting triplets, 19 \% in the isolated CIG [7], and 23 \% in the Northern and Southern triplets. This suggests that the maximum excess of early-type galaxies in triplets compared to isolated galaxies is observed in the sample of Interacting triplets, which are the ''oldest'' systems.

Our results lead us to conclude that the hypothesis of dark matter concentrated in the halos of indi­vidual galaxies is better suited for dynamically ''young'' LS triplets. Note that we reached a similar conclu­sion previously [10] by analyzing the virial and total masses of galaxy triplets. The hypothesis about dark matter concentrated in the common halo of a triplet is consistent with compact systems, which the Inter­acting systems and some of the Northern and Southern systems are.

\bigskip

{\bf ACKNOWLEDGMENTS.} We wish to thank V.E. Karachentseva for helpful comments to the paper. This work was supported by the State Foundation for Basic Research of the Ministry of Education and Science of Ukraine (grant no. F7/267-2001). We used the LEDA database (http://leda.univ-lyon1.fr) and NED (http://nedwww.ipac.caltech.edu). 

{}

\begin{thebibliography}{}

\bibitem {} T. A. Agekyan and Zh. P. Anosova. Astron. Zh. 44, 1261 (1967).

\bibitem {} Zh P. Anosova, Astrofizika 27, 535 (1987).

\bibitem {} Zh. P. Anosova, A. V. Ivanov, L. G. Kiseleva el al., Astron. Zh. 70, 943 (1993).

\bibitem {} I. B. Vavilova, V. E. Karachentseva, D. I. Makarov, O. V. Melnyk, Kin. Phys. Celest. Bodies 21, 1 (2005), astro-ph/0609622.

\bibitem {}  A. V. Ivanov, L. A. Filislov, A. D. Chernin, Astron. Zh. 72, 416 (1995).

\bibitem {} A. V. Ivanov, A. D. Chernin, Pis'ma Astron. Zh. 17, 569 (1991).

\bibitem {} V. E. Karachentseva, Soobshch. Spets. Astrofiz. Obs. No. 8, 3 (1973).

\bibitem {} V. E. Karachentseva, I. D. Karachentsev, Astron. Zh. 77, 569 (2000).

\bibitem {} V. E. Karachentseva, I. D. Karachentsev, V. S. Lebedev, Izv. Spets. Astrofiz. Obs. 26, 42 (1987).

\bibitem {} V. E. Karachentseva, O. V. Melnyk, I. B. Vavilova, D. I. Makarov, Kin. Phys. Celest. Bodies 21, 217 (2005).

\bibitem {}  L.S. Kiseleva, Kin. Fiz. Nebesn. Tel 3, 67 (1987).

\bibitem {}  L.S. Kiseleva, V. V. Orlov, Soobshch. Spets. Astrofiz. Obs. No. 60, 23 (1989).

\bibitem {} L.S. Kiseleva, A  D. Chernin, Pis'ma Astron. Zh. 14, 970 (1988).

\bibitem {}  L.S. Kiseleva, A  D. Chernin,  Soobshch. Spets. Astrofiz. Obs. No. 60, 5 (1989).

\bibitem {}  O. V. Melnyk, Pis'ma Astron. Zh. 32, 302 (2006), astro-ph/0612353.

\bibitem {}  A. V. Trofimov, A. D. Chernin, Astron. Zh. 72, 308 (1995).

\bibitem {}  A  D. Chernin, A. V. Trofimov, A. V. Ivanov, Astron. Tsirk. No. 1540, 3 (1989).

\bibitem {}  H. Aceves, Mon. Not. R. Astron. Soc. 326, 1412 (2002).

\bibitem {}  A  D. Chernin, A. V. Ivanov, A. V. Trofimov, S. Mikkola, Astron. Astrophys. 281, 685 (1994).

\bibitem {}  R. Coziol, E. Brinks, H. Bravo-Alfaro, Astron. J. 128, 68 (2004).

\bibitem {} C. Da Rocha, C. M. de Oliveira, B. L. Ziegler, EAS Publ. Ser. 20, 273 (2006).

\bibitem {} M.J.Geller, J.P Huchra, Astrophys. J. Suppl. Ser. 52, 61 (1983).

\bibitem {} P. Hickson,.E. Kindl, J. R. Auman, Astrophys. J., Suppl. Ser. 70, 687 (1989).

\bibitem {}  Z. Y. Huo, X Y. Xia, S. J. Xue et al., Astrophys. J. 611, 208 (2004).

\bibitem {} L.S. Kiseleva, ASP Conf. Ser. 209, 388 (2000).

\bibitem {}  M. A. G. Maia, L. N. da Costa, D. W. Latham, Astrophys. J. Suppl. Ser. 69, 809 (1989).

\bibitem {}  J. S. Mulchaey, D. S. Davis, R. F. Mushotzky, D. Burstein, Astrophys. J.  Suppl. Ser. 145, 39 (2003).

\bibitem {}  M. Valtonen, S. Mikkola, Ann. Rev. Astron. Astrophys. 29, 9 (1991).

\bibitem {} B. A. Vorontsov-Velyaminov, R. I. Noskova, V. P. Arkhipova, Astron. Astrophys. Trans. 20, 717 (2001).

\bibitem {} J.-Q. Zheng, M. Valtonen, A. D. Chernin, Astron. J. 106, 2047 (1993).

\end{thebibliography}
\end{document}